# Sliding Mode Network Perimeter Control

Youssef Bichiou, Maha Elouni, Hossam M. Abdelghaffar and Hesham A. Rakha, *Fellow, IEEE*

*Abstract*—Urban traffic congestion is a chronic problem faced by many cities in the US and worldwide. It results in inefficient infrastructure use as well as increased vehicle fuel consumption and emission levels. Congestion is intertwined with delay, as road users waste precious hours on the road, which in turn reduces productivity. Researchers have developed, and continue to develop, tools and systems to alleviate this problem. Network perimeter control is one such tool that has been studied extensively. It attempts to control the flow of vehicles entering a protected area to ensure that the congested regime predetermined by the Network Fundamental Diagram (NFD) is not reached. In this paper, an approach derived from sliding mode control theory is presented. Its main advantages over proportional-integral controllers include (1) minimal tuning, (2) no linearization of the governing equations, (3) no assumptions with regard to the shape of the NFD, and (4) ability to handle various demand profiles without the need to retune the controller. A sliding mode controller was implemented and tested on a congested grid network. The results show that the proposed controller produces network-wide delay savings and disperses congestion effectively.

## I. INTRODUCTION

ROADWAY traffic is a complex phenomenon that requires serious modeling scalability. Nevertheless, various properties can be directly observed. These properties include (1) traffic stream density ($k$): the number of vehicles per unit length of the road or lane; (2) traffic stream flow ($q$): the number of vehicles passing a fixed point per unit of time; and (3) space-mean speed ($u$): the traffic stream density weighted average speed. These three parameters are related using the hydrodynamic equation $q = ku$. Furthermore, it has been hypothesized that these variables along a roadway segment are related to one another, forming what is commonly known as the fundamental diagram [1, 2]. In addition, a number of states can be identified on the fundamental diagram. These include the free-flow speed ($u_f$), which is the traffic stream space-mean speed when the roadway is empty, and the speed-at-capacity ($u_c$), which is the traffic stream space-mean when the flow is at maximum (prior to the onset of congestion).

These are the key properties used to model the behavior of vehicles in any given network. The flow continuity equation

(i.e., conservation of mass) demonstrates that the flow through a network is directly related to the density on the different roads. This relationship is characterized by the Network Fundamental Diagram (NFD), as presented in Fig. 1. If the distance-weighted arithmetic means of the link densities and flows are computed, a scaled NFD can be obtained. In general, obtaining a clear and well-behaved NFD can be difficult. The scatter of points (as presented in Fig. 1) might not generate a visible curve. However, if the distribution of congestion is uniform across the network, the existence of a well-defined NFD is guaranteed [3] [4] [5]. Godfrey *et al*. [6] were the first to introduce the concept of the NFD for the center of London. The authors demonstrated that the relationship between the average velocity and the vehicle-travelled distance is parabolic and that the speed is inversely proportional to the density. Geroliminis *et al*. [7] observed the NFD in the congested urban network of Yokohama, Japan, demonstrating that under homogeneous conditions, even with large discrepancies between the different link fundamental diagrams, the NFD of the network can be reproduced. This means that the NFD is a direct result of the infrastructure and thus is independent of the number of vehicles in circulation, the demand, and the selected routes (i.e., independent of the road taken by each individual vehicle and the origin-destination table).

Leclercq *et al*. [8] acknowledges that the shape and scatter of the NFD is subject to local traffic heterogeneities. This is achieved through a non-homogeneous distribution of congestion due to heterogeneous local capacities and route choice. He also concludes that it is difficult to link and understand the connection between local phenomena and the NFD.

Many clustering algorithms have been developed in order to determine small regions where there is small variance in density and an NFD can be generated [9-15]. With the determination of multi-region networks, researchers developed multi-region controllers [16-18].

Other studies focused on controlling only one region network. In order to avoid congestion in a specific region of a given network (i.e., a protected network [PN]), flow optimization [19] can be useful. However, we will focus on gating, also referred to as perimeter control. There are gating solutions to reduce congestion, travel time, and delay, some of which are based on the NFD and others that are not [20-28]. This paper focuses on methods that rely, in one form or another,

This effort is funded by the Department of Energy through the Office of Energy Efficiency and Renewable Energy (EERE), Vehicle Technologies Office, Energy Efficient Mobility Systems Program under award number DE-EE0008209 [Product # DOE-VT-0008209-J05].

Youssef Bichiou is with the Center for Sustainable Mobility, Virginia Tech Transportation Institute, 3500 Transportation Research Plaza, Virginia Tech, Blacksburg, VA 24061 USA (e-mail: ybichiou@vtti.vt.edu).

Maha Elouni is with the Center for Sustainable Mobility, Virginia Tech Transportation Institute, 3500 Transportation Research Plaza, Virginia Tech, Blacksburg, VA 24061 USA (e-mail: emaha@vt.edu).

Hossam M. Abdelghaffar is with the Department of Computer Engineering and Systems, Faculty of Engineering, Mansoura University, Mansoura 35516, Egypt, and also with the Center for Sustainable Mobility, Virginia Tech Transportation Institute, Virginia Tech, Blacksburg, VA 24061 USA (e-mail: hossam_wahed@mans.edu.eg).

Hesham A. Rakha is with the Charles E. Via, Jr. Department of Civil and Environmental Engineering, Virginia Tech, Blacksburg, VA 24061 USA, and also with the Center for Sustainable Mobility, Virginia Tech Transportation Institute, Virginia Tech, Blacksburg, VA 24061 USA (e-mail: hrakha@vt.edu).

on the NFD. The idea behind these methods is to control the access points to a "protected" area (i.e., protected subnetwork) by regulating the entering traffic to ensure that the network will not operate beyond the capacity regime (i.e., maximum throughput). This can be achieved given information provided by the NFD. The aim is to avoid oversaturation (i.e., the congested regime, as shown in Fig. 1).

Various researchers have tried to achieve this objective [29-36]. Li *et al.* [37] investigated a perimeter control strategy for an oversaturated network. They optimized the green duration allocation in order to maximize the throughput using a genetic algorithm to minimize queues and delays by optimizing phase sequences and offsets. However, their method used fixed signal timings, which does not reflect typical real-time traffic conditions [29-36].

Many studies have overcome this issue and achieved real-time perimeter control using techniques such as the standard proportional-integral controller (PIC) [38-41], a robust PIC [10], a linear quadratic controller [14], and a model predictive controller. Haddad *et al.* [30-32] introduced various adaptive perimeter controller schemes that take into account model uncertainty and NFD scatter. However, due to the nonlinear nature of the NFD, model linearization is essential for controller design.

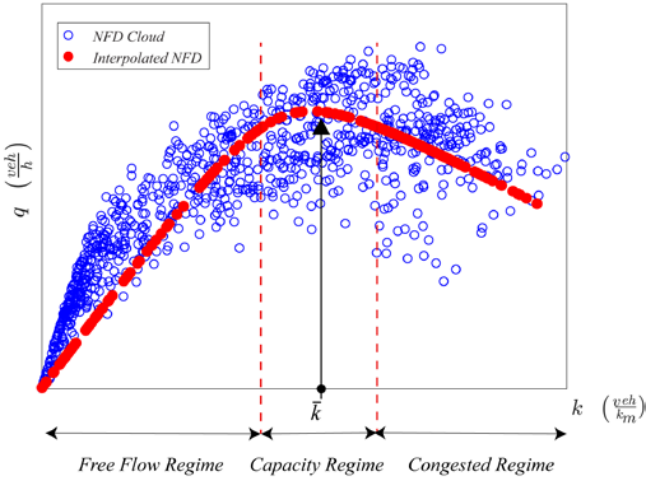

Fig. 1. The different segments of a typical NFD.

The PIC requires a number of input parameters, namely: a set point (the desired network density) and two gains that have to be tuned a priori. The optimum gain values are typically difficult to compute using the various methods to estimate the best gains [42-44]. Anandanatarajan *et al.* [45] and Sung *et al.* [46] showed further limitations of PI and proportional-integral-derivative (PID) controllers. Elouni *et al.* [47] demonstrated that using a weather-tuned perimeter control, (i.e., retuning the PIC gains with respect to weather), improved the network's performance metrics in terms of average speed and delay. Kouvelas *et al.* developed robust adaptive tuning techniques that alleviated the tuning problem [48]. In addition, the PIC developed in [38] requires linearization of the control function, which further complicates and limits its use.

Mirkin *et al.* [49] developed an adaptive sliding mode controller (SMC) that takes into account uncertainties and unknowns in the NFD, delayed input, and adjustable gains.

However, it requires numerous design parameters and needs model linearization.

In this work, we developed an SMC [50] that has comparable performance to the PIC, avoids the need for tuning, and makes no assumptions about the governing model (i.e., no linearization is needed for the NFD). The present effort shows that only a set point (i.e., a target network vehicle density) is needed. This value is obtained only once from the NFD. The other parameters for this new controller can be evaluated in a systematic manner.

This paper is composed of seven sections. Section II presents the NFD equations. Section III is a brief description of the PIC used in the literature. The introduced SMC is presented in Section IV. Section V presents a case study to evaluate both controllers. The results are discussed in Section VI, followed by concluding remarks.

## II. DERIVATION OF THE NFD EQUATIONS

NFD equations are used to plot the NFD curve in order to determine the set point for the SMC and the PIC used for comparison purposes. It has to be noted that NFD equations are used in the PIC modelling but not in the SMC.

The NFD is computed based on the average link density ($k$) in vehicles per unit length and the average vehicle flow ($q$) inside the network in vehicles per unit time. These quantities can be computed from loop detectors placed throughout the network, where $k$ is computed using (1).

$$k[n] = \frac{1}{L} \sum_{z \in Z} k_z[n].\, l_z \qquad (1)$$

where $z$ is the index for the link; $Z$ is the set of all links belonging to the protected area where measurements are taking place; $n$ is the time index and reflects the cycle number; $L$ is the total length of the PN (i.e., the sum of the length of all links $L = \sum_{z \in Z} l_z$; these links also feature loop detectors.); $l_z$ is the length of link $z$; and $k_z[n]$ is the traffic stream density on link $z$ during cycle $n$ and is calculated using (2).

$$k_z[n] = n l_z \, k_{j\,z} \frac{o_z[n-1]}{100} \qquad (2)$$

where $n l_z$ is the number of lanes of link $z$; $k_{j\,z}$ is the jam density of link $z$; $o_z$ is the measured time-occupancy (in percent) on link $z$.

The flow inside the network ($q$) is calculated using (3):

$$q[n] = \frac{1}{L} \sum_{z \in Z} q_z[n].\, l_z \qquad (3)$$

where $q_z[n]$ is the measured flow on link $z$.

The NFD is the plot relating the flow $q[n]$ to the network density $k[n]$.

$$q[n] = \varphi(k[n]) \qquad (4)$$

where $\varphi$ is an unknown NFD function.

It is important to mention here that knowledge of the exact $\varphi$ function is not necessary. The plot of the NFD scatter (measurements of $q$ and k during the simulation) is used to determine the set point that corresponds to the density having the highest flow.

## III. Proportional Integral Controller (PIC)

This section introduces the PIC. Any given network can be represented by an abstract schematic (Fig. 2). This network experiences moderate to heavy congestion during the course of the day, leading to delays, excessive fuel consumption, and pollution. To alleviate these effects, the identified area of congestion (PN) is protected. That is, all the inflows (i.e., $q_{in}^1$, $q_{in}^2$, etc.) are monitored and in some cases restricted to the extent that the protected area operates at capacity.

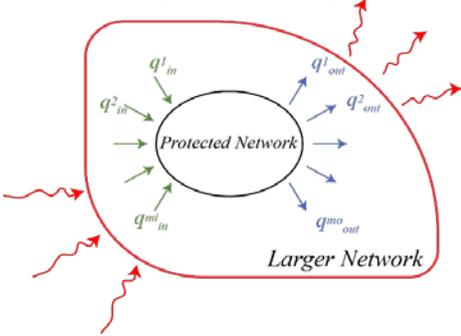

Fig. 2. An abstract schematic of an urban network that features a congestion protected subnetwork.

The time rate of change of the number of vehicles inside the PN is computed using (5).

$$\frac{dNV(t)}{dt} = q_{in}(t) - q_{out}(t) + q_d(t) \qquad (5)$$

where $q_{in}(t) = \sum_{i=1}^{mi} q_{in}^i(t)$ is the sum of all inflows at the instant $t$; $q_{out}(t) = \sum_{i=1}^{mo} q_{out}^i(t)$ is the sum of all vehicle outflows from the subnetwork; $q_d(t)$ is the disturbance flow that might occur inside the protected area; and $NV = k \times L$ is the number of vehicles inside the PN. $NV$ is the product of the vehicle density $k$ and the total length of the links $L$ inside the protected region.

In this section, we describe a few details of the PIC used in the literature [38, 40]. Since the objective is to avoid congestion, a desired density $\bar{k}$ is sought within a PN. This density is usually chosen to be the density at capacity (Fig. 1). Therefore, the error defined in (6) is driven to zero when the control is activated.

$$e(t) = k(t) - \bar{k} \qquad (6)$$

where $k$ is the current density of vehicles in the protected area, and $\bar{k}$ is the desired density (density at capacity).

The PIC equation for discrete time steps is given by [40] as

$$q_{in}[n] = q_{in}[n-1] - K_p(k[n] - k[n-1]) \qquad (7)$$
$$+ K_I(\bar{k} - k[n])$$

where $n$ is the time index and $K_p = \mu/\zeta$ and $K_I = (1-\mu)/\zeta$ [38].

Based on measured data and the least squares method, $\mu$ and $\zeta$ are estimated using (8) [40].

$$k[n+1] - \bar{k} = \mu \cdot (k[n] - \bar{k}) + \zeta \cdot (q_{in}[n] - \bar{q}_{in}) \qquad (8)$$
$$+ \varepsilon[n]$$

where $\bar{q}_{in}$ is the average measured inflow of vehicles at the capacity regime and $\varepsilon[n]$ is an error term.

The values of $\mu$ and $\zeta$ are determined using data from the NFD (i.e., scatter points) in the capacity regime [40]. Tuning, in practice, could deliver negative and/or zero values. Particular care must be taken with respect to the calibration data so that appropriate $\mu$ and $\zeta$ values are obtained. There also exist some robust tuning techniques that alleviate this problem [48].

During simulation, the values of density $k$ at the previous time step, as calculated using loop detectors, and the current time step are needed. These are used to calculate the flow that should enter the PN while avoiding congestion.

In general, the tuning process and the estimation of the density ($k$) add a layer of complexity to the controller. Consequently, developing an approach that does not require tuning, does not require linearization of the control equation, and provides robustness is of interest.

## IV. Proposed Sliding Mode Controller (SMC)

This section introduces the developed SMC. For this purpose, we simplify the notation. Equation (5) can be re-written as

$$\frac{dk(t)}{dt} = \frac{q_{in}(t) - q_{out}(t) + q_d(t)}{L} \qquad (9)$$
$$= u(t) - V_{out}(t) + V_d(t)$$

where we assume that $u = q_{in}/L$ is the system's input. This input governs the maximum number of vehicles allowed to enter the protected area at instant $t$ (i.e., $q_{in}$). For discrete time steps, $u[n]$ (i.e., $n$ is the corresponding time step) would determine the maximum number of vehicles allowed over a given time horizon $\Delta t$. Equation (9) governs the time rate of change of the density ($k$) in the protected area. Here $k$ is the state variable used in this work and represents the vehicle density inside the protected area. Specifically, it is the length weighted average network density computed as the sum of the product of the link densities and lengths divided by the total length of the links in the protected area.

It is important to note here that $u(t)$ is proportional to the sum of all inflows entering the protected area through the $mi$ access points (i.e. access links).

$$u(t) = \frac{q_{in}(t)}{L} = \frac{1}{L} \sum_{i=1}^{mi} q_{in}^i(t) \qquad (10)$$

To derive an expression for this input, we use sliding mode control theory [51] and introduce the error $e$ as $e = k - \bar{k}$, where $\bar{k}$ is a set point around which the PN is desired to operate. In this case, this point corresponds to the density at capacity.

We introduce the variable $x$ defined as

$$x(t) = \int_0^t e(\tau) \, d\tau = \int_0^t (k(\tau) - \bar{k}) \, d\tau \qquad (11)$$

and $S$ as

$$S = \frac{dx(t)}{dt} + \lambda x(t) = \dot{x}(t) + \lambda x(t) \qquad (12)$$

where $\lambda$ is a strictly positive real number.

The sliding surface $S$ is defined in Equation (13). This leads to the relationship $\dot{x}(t) + \lambda x(t) = 0$. In other words, $x$ decays exponentially to zero given that $\lambda$ is strictly positive.

$$S(x(t)) = 0 \qquad (13)$$

Using (9), (11), and (12), we obtain

$$\frac{dS(t)}{dt} = u(t) - V_{out}(t) + V_d(t) + \lambda(k(t) - \bar{k}) \qquad (14)$$

Since the trajectories are expected to remain on the surface (i.e., (13)) for all time, $dS(t)/dt$ has to remain at zero (i.e., $dS(t)/dt = 0$). This in turn leads to (15).

$$u^*(t) = V_{out}(t) - V_d(t) - \lambda\big(k(t) - \bar{k}\big) \qquad (15)$$

The values of $V_{out}(t)$ and $V_d(t)$ are not known since they represent the current outflow and disturbance flow in the network. Consequently, we use bounded estimates $\hat{V}_{out}$ and $\hat{V}_d$ with

$$\begin{aligned}\big|\hat{V}_{out}(t) - V_{out}(t)\big| &\leq \alpha \\ \big|\hat{V}_d(t) - V_d(t)\big| &\leq \beta\end{aligned} \qquad (16)$$

where $\alpha$ and $\beta$ are positive real numbers. It is important to note here that the bounds might not be always available or known. In this case, adaptive SMC can be used. In this approach, the control gains are adapted dynamically to counteract the uncertainties. For further details, the reader is referred to the work of Utkin et al. [52]. In this effort, we assume that $\alpha$ and $\beta$ can be determined.

Using the previous estimates, the new estimated controller $\hat{u}$ becomes

$$\hat{u}(t) = \hat{V}_{out}(t) - \hat{V}_d(t) - \lambda\big(k(t) - \bar{k}\big) \qquad (17)$$

It is important to note here that we chose our estimates to be in the following form: $\hat{V}_{out}(t) = V_{out}(t - \Delta t)$ and $\hat{V}_d(t) = V_d(t - \Delta t)$. That is, our estimates are the observed measurements of the mentioned quantities at the previous time step. This assumption requires loop detectors placed at the exits and on links in the PN.

We add to the estimated controller the term $\gamma\, \text{sign}(S)$ as

$$u_s = \hat{u} - \gamma\, \text{sign}(S), \qquad (18)$$

where $\gamma$ is a positive real number, to force the controller to always move toward the sliding surface.

That leads to

$$\begin{aligned}u_s = \hat{V}_{out} - \hat{V}_d &- \lambda\big(k - \bar{k}\big) \\ &- \gamma\, sign\left(k - \bar{k}\right. \\ &\left. - \lambda\int_0^t k - \bar{k}\, dt\right)\end{aligned} \qquad (19)$$

For a discrete time step $n$,

$$\begin{aligned}u_s[n] = \hat{V}_{out}[n] - \hat{V}_d[n] &- \lambda\big(k[n] - \bar{k}\big) \\ &- \gamma\, sign\left(k[n] - \bar{k}\right. \\ &\left. - \lambda\sum_{i=0}^{n}\big(k[i] - \bar{k}\big)\Delta t\right)\end{aligned} \qquad (20)$$

To ensure the surface defined by (13) is a stable surface for the chosen controller (19), we introduce the Lyapunov function, defined by $LF(S)$.

$$LF(S(x)) = \frac{1}{2}S(x)^T S(x) = \frac{1}{2}\|S(X)\|^2 \qquad (21)$$

The equilibrium (13) is stable if

$$\frac{d}{dt}\big(\,LF(S)\,\big) \leq 0 \qquad (22)$$

At this point it is assumed that the actual control $u(t)$ input to the system is bounded:

$$u(t) = \begin{cases} U_{max} & if\ u_s \in [U_{max}, \infty[ \\ u_s & if\ u_s \in [U_{min}, U_{max}] \\ U_{min} & if\ u_s \in ]-\infty, U_{min}] \end{cases} \qquad (23)$$

Therefore, using Equation (14) we derive

$$\begin{aligned}\frac{d}{dt}\big(\,LF(S)\,\big) = S\big(u(t) - u_s(t)\big) &+ S\,u_s(t) \\ &- S\,V_{out}(t) + S\,V_d(t) \\ &+ \lambda\,S\big(k(t) - \bar{k}\big)\end{aligned} \qquad (24)$$

If $u(t)$ does not hit the bounds, then $u(t) = u_s(t)$. Using Equations (17) and (18) we compute

$$\begin{aligned}\frac{d}{dt}\big(\,LF(S)\,\big) = S\,&\left(\hat{V}_{out} - V_{out}(t)\right) \\ &+ S\left(V_d(t) - \hat{V}_d(t)\right) - \gamma|S|\end{aligned} \qquad (25)$$

Using Equation (16) we obtain

$$\frac{d}{dt}\big(\,LF(S)\,\big) \leq \alpha\,|S| + \beta|S| - \gamma|S| \qquad (26)$$

Choosing $\gamma = \alpha + \beta + \eta$ (with $\eta$ strictly positive) leads to

$$\frac{dLF(S)}{dt} \leq -\eta\,|S| \qquad (27)$$

If $u(t)$ does hit the bounds, then $u(t) = U_b$ where $U_b$ refers to any of the bounds. In this case using Equations (17) and (18)

$$\begin{aligned}\frac{d}{dt}\big(\,LF(S)\,\big) = S(U_b - u_s) \\ + S\left(\hat{V}_{out} - V_{out}(t)\right) \\ + S\left(V_d(t) - \hat{V}_d(t)\right) - \gamma|S|\end{aligned} \qquad (28)$$

Using Equation (16) we obtain

$$\frac{d}{dt}\big(\,LF(S)\,\big) \leq |S|\,|U_b - u_s| + \alpha\,|S| + \beta|S| - \gamma|S| \qquad (29)$$

There exists a positive real number $\theta$ such that $\theta > |U_b - u_s|$.

Choosing $\gamma = \alpha + \beta + \theta + \eta$ (with $\eta$ strictly positive) leads to

$$\frac{dLF(S)}{dt} \leq -\eta\,|S|. \qquad (30)$$

In other words, the distance to the sliding surface decreases with time. This sliding surface $S = \dot{x} + \lambda\,x = 0$ will be reached in a finite time that is bounded by $|S(t = 0)/\eta|$. Consequently, the choice of $\eta$ will impact the time it takes the system to reach the sliding surface. Once on the surface, the system will remain there and will slide to the desired state, which in this case is $x = 0$ exponentially with a time constant $\lambda^{-1}$.

It is important to note here that the value of $k[n]$ is unknown at the current time step $n$. Therefore, quantities measured from the previous time steps were used as shown in equation (31).

$$\begin{aligned}\hat{V}_{out}[n] &= q_{out}[n-1]/L \\ \hat{V}_d[n] &= q_d[n-1]/L\end{aligned} \qquad (31)$$

Also, the value of $k[n]$ was replaced with $k[n - 1]$. Hence, the controller becomes

$$\begin{aligned}u[n] = \frac{q_{out}[n-1]}{L} &- \frac{q_d[n-1]}{L} \\ &- \lambda\big(k[n-1] - \bar{k}\big) \\ &- (\alpha + \beta + \eta)\, sign\left(k[n-1] - \bar{k}\right. \\ &\left. - \lambda\sum_{i=0}^{n-1}\big(k[i] - \bar{k}\big)\Delta t\right)\end{aligned} \qquad (32)$$

It is important to note here that the SMC suffers from a known problem—control/input chattering. This can be solved by changing the "sign()" function in (18) into a saturation "sat()" function or a "tanh()" function. Chattering usually results in the failure of components in mechanical systems. However, in this research, the values of the inflow allowed in the protected area are transformed into green times (explained later in the paper; (34)). For each new cycle $\Delta t$ (assumed to be 60 s in this paper) a new value of the green time is obtained. Chattering in this case will result in the change in the green time by a large or small value from one cycle to the next. This change is expected not to cause a failure of any component and thus is less of an issue in this application. In contrast to mechanical systems, chattering (i.e., sudden change of the control input) might cause the failure of components, for instance a sudden variation of the velocity input signal to a motor might damage it.

For the controller to function in an ideal manner, detectors are needed on the majority of the links of the protected area as well as at the entrance points. This is also true for the PIC to monitor the state of the area in terms of associated vehicle density, inflow, and outflow. This information can then be used to determine the estimated values needed by the controller (i.e., $\hat{V}_{out}(t)$ and $\hat{V}_d(t)$ (31)). The developed controller features four hyper-parameters, namely $\alpha, \beta, \lambda$, and $\eta$. $\alpha$ and $\beta$ quantify how close the estimated outflow and disturbance flows are to the actual values at a particular instant in time. In the next section, this control law will be implemented and applied to a grid network.

## V. CASE STUDY

In this section, the introduced grid network is used to conduct the controllers testing (Fig. 3). The network was created to replicate downtown Washington, DC, both in terms of its one-way streets and block lengths. This network includes a protected region, also referred to as the PN. The PN corresponds to the region surrounded by the green rectangle (Fig. 3). All the access points to this subnetwork are identified with the yellow arrows. In total there are eight links that feed directly into the protected area. For the purpose of this paper, $u_{min} = 480$ veh/h and $u_{max} = 12,960$ veh/h. These bounds represent the minimum and maximum allowable vehicle flow into the protected area via those eight links (Fig. 3).

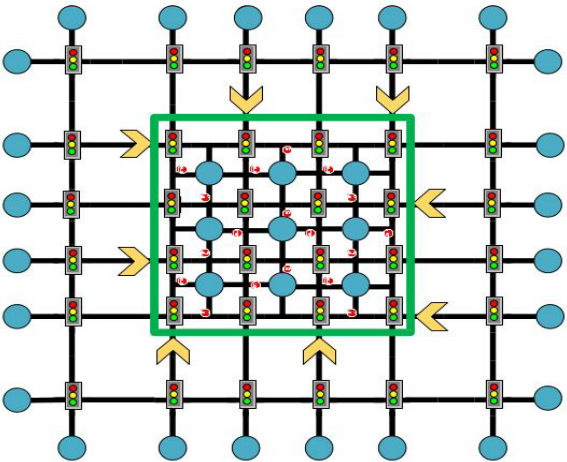

Fig. 3. Grid network modeled in INTEGRATION.

The PN contains 48 links, all of which are one-way roadways and each of which has only one lane of the same length of 150 m. The full network contains 36 signalized intersections. The network was modeled using INTEGRATION software. The computation of various measures of effectiveness within

INTEGRATION is beyond the scope of this paper. The delays were validated in [53], the computation of vehicle stops was validated in [54], and the estimation of vehicle fuel consumption and emission levels are computed using the VT-Micro model, which has also been extensively validated [55-57]. Origin and destination zones for trips are represented by blue circles. Loop detectors are placed on each link of the network and collect measurements every cycle. The cycle length is taken to be 60 s (i.e., $\Delta t = 60 \, s$). The demand for this network (D1) is generated during the first 75 minutes of the simulation, increasing during the first 37.5 minutes and decreasing after that to model the buildup and decay of traffic congestion (Fig. 4). The simulation time is taken to be 176 minutes in order to provide sufficient time for all vehicles to clear the network. Dynamic traffic assignment was activated during the simulation to reflect realistic driver behavior during congested conditions (i.e., rerouting of vehicles is activated).

The NFD associated with the network presented in Fig. 3 is shown in Fig. 5. During all "no control" case simulations, the green time for the different traffic signals was optimized using phase split optimization (Webster method [58]). The offset is also optimized using the procedures described in [59]. The point cloud, shown in Fig. 5, presents snapshots of the simulated sub-network state. Note that congestion is observed beyond an average network density of $\bar{k} = 48$ veh/km. Beyond this point, vehicles experience significant delays and consume additional fuel. To avoid reaching this state, the proposed SMC was implemented as described earlier in Section IV.

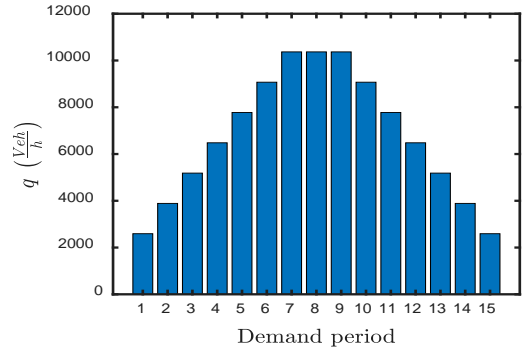

Fig. 4. Base demand profile used in the simulation where each demand period spans 300 s. This demand profile is referred to as demand D1 in the text.

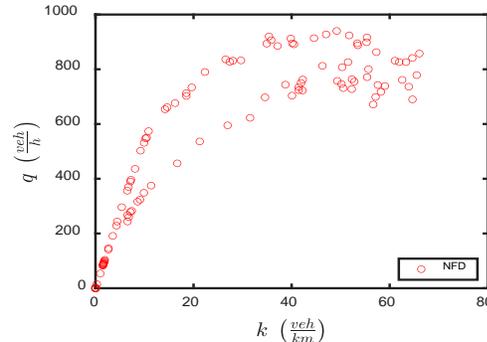

Fig. 5. NFD of the grid network presented in Fig. 3 for the demand profile presented in Fig. 4.

## VI. RESULTS AND SENSITIVITY ANALYSIS

This section describes the various numerical simulations that were performed. In the first subsection, the response of the SMC is shown and compared to the PIC for various controller parameters. The inflow and outflow from the protected area for the case of No Perimeter Control (NPC) and SMC are presented in the second subsection. In the third subsection, the behavior of the SMC for a range of uncertainty is presented. Finally, the performance of the SMC and PIC with respect to various demand profiles is shown.

### A. Performance of the SMC and PIC for Different Parameters

Tuning is required for the PIC. This requirement is not needed for the SMC. Only two parameters, $\lambda$ and $\eta$, are considered. $\eta$ governs how fast the system converges to the sliding surface. $|S(x(t=0))/\eta|$ is the maximum time it takes for convergence to occur. Since we choose the control to be activated when $|(k-\bar{k})/k| \leq 0.15$ (i.e., control is activated when $k$ reaches $0.85\ \bar{k}$), the value of $|S(x(t=0))/\eta|$ is small for a wide range of $\eta$.

$$\left| \frac{S(x(t=0))}{\eta} \right| \leq 0.15\ \frac{\bar{k}}{\eta} \qquad (33)$$

Once on the surface (i.e., $\dot{x} + \lambda\ x = 0$), the system converges exponentially to zero with a time constant of $\lambda^{-1}$ with the requirement that $\lambda$ must be less than the minimum frequency of the un-modeled dynamics of the system [51]. This means that we can choose $\lambda$ as small as adequately possible. From the previous results, we can conclude that almost no tuning is needed for the parameters of the SMC. The model presents only ranges of validity.

Table 1 presents various responses of the system using different control parameters for the PIC and SMC for the traffic demand presented in Fig. 4. The table also presents the base case (i.e., NPC with optimum signal timing for each intersection) relative to which all comparisons are performed. A negative change means a decrease in value, while a positive change indicates an increase in value. For the SMC, we tested various combinations, in all cases we get a reduction with respect to the base case. Fig. 6 and Fig. 7 present a comparison between the evolution of the density $k$ inside the protected area in the presence and absence of control. In Fig. 6, the five tuned cases of the PIC were picked, and in Fig. 7 the best performing SMC cases in terms of travel time reduction were picked. It is important to mention here that for the SMC the density in the protected area is generally below the target density $\bar{k}$ with the exception of SMC 6, which is slightly above $\bar{k}$ for the first few time steps when the control is activated. PIC 1 and 2 are clearly above the target value by a significant margin and over the entire duration of the control interval (i.e., the time steps where the density is above $0.85\ \bar{k}$). This is clear for PIC 1, which exceeds the NPC case in some time steps (i.e., congestion forms in the protected area). This in turn underscores the criticality of the tuning process and its importance in the PIC performance.

In the rest of the paper, and since the performance of most of the various controllers with different parameters is almost the same with respect to the reduction in travel time (Table 1), the parameters of SMC 6 and PIC 5, which deliver the highest reduction in delay, will be used.

| | | | TT(s) | Delay(s) | Fuel(l) | Speed (km/h) |
|---|---|---|---|---|---|---|
| **Base Case (NPC)** | | | 659.26 | 247.28 | 0.42 | 14.97 |
| **SMC** | | | | | | |
| **SMC** | $\lambda$ | $\eta$ | **Change (%)** | **Change (%)** | **Change (%)** | **Change (%)** |
| 1 | 1.5 | 2.0 | -12.31 | -12.88 | -4.63 | 11.86 |
| 2 | 1.5 | 20.0 | -12.92 | -15.57 | -5.75 | 12.06 |
| 3 | 1.5 | 200.0 | -8.33 | -7.90 | -2.65 | 7.93 |
| 4 | 15.0 | 2.0 | -13.19 | -14.56 | -4.77 | 13.46 |
| 5 | 15.0 | 20.0 | -12.61 | -15.82 | -5.41 | 12.23 |
| 6 | 15.0 | 200.0 | -12.56 | -19.51 | -7.11 | 11.30 |
| 7 | 150.0 | 2.0 | -8.99 | -12.84 | -4.58 | 7.90 |
| 8 | 150.0 | 20.0 | -7.90 | -13.24 | -4.48 | 7.00 |
| 9 | 150.0 | 200.0 | -4.62 | -6.08 | -2.07 | 4.02 |
| **PIC** | | | | | | |
| **PIC** | $\mu$ | $\zeta$ | **Change (%)** | **Change (%)** | **Change (%)** | **Change (%)** |
| 1 | 0.801 | 0.002 | -10.63 | -14.75 | -5.294 | 9.63 |
| 2 | 0.547 | 0.002 | -8.451 | -10.54 | -3.506 | 8.08 |
| 3 | 0.885 | 0.003 | -12.5 | -16.1 | -5.7 | 11.76 |
| 4 | 0.747 | 0.003 | -11.53 | -15.36 | -5.29 | 11.03 |
| 5 | 0.847 | 0.002 | -12.95 | -17.52 | -6.25 | 12.12 |

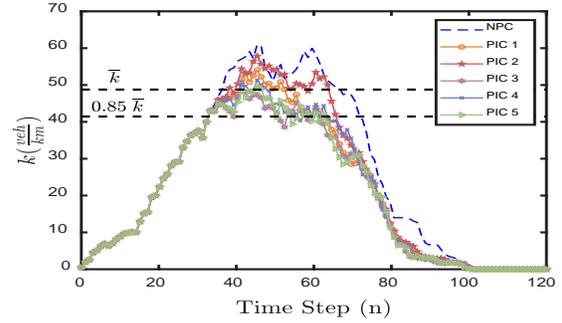

Fig. 6. Time series of the density $k$ from the protected area when the PIC is activated and using the tuned parameters.

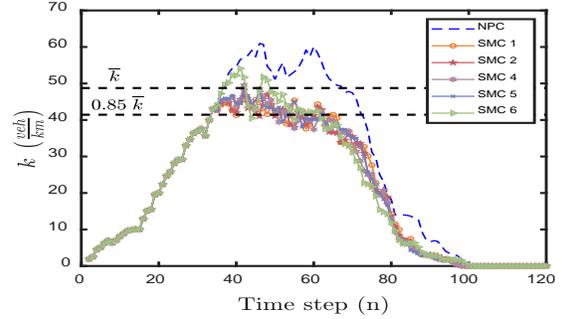

Fig. 7. Time series of the density $k$ from the protected area when the SMC control is activated. The five best cases are shown.

Fig. 8 shows a comparison between two controllers: the proposed SMC and the PIC for the demand profile D1. The plot shows similar performance between the two controllers. The tuned parameters for the PIC are taken to be $\mu = 0.847$ and $\zeta = 0.002$ (i.e., $K_I = 73.7$, $K_P = 408$), and $\bar{k} = 48.76$ veh/km). For the SMC, the parameters are $\lambda = 15$ and $\eta = 200$.

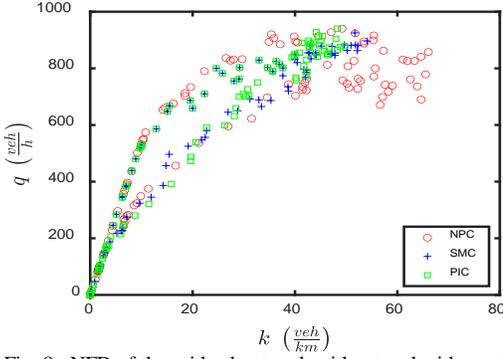

Fig. 8. NFD of the grid subnetwork without and with control.

In the absence of control, we notice that the average network density of the protected area exceeds the optimum value $\bar{k}$. Once the control is activated, congestion is consistently eliminated from the protected area.

### B. Inflow and Outflow Profiles With and Without SMC

Fig. 9 presents the time series of the inflow of vehicles without perimeter control (NPC) and with perimeter control for the demand profile D1 presented in Fig. 4. It is important to mention here that inflow computed by the controller ($q_{in}$) is transformed into green times ($G$) that are allocated to the different approaches feeding into the protected area. This is performed using Equation (34). This transformation creates discrepancies between the computed control inflow and the actual inflow when the control is activated (Fig. 9), given that the effective green time is different from the displayed green time because of start loss and end gain (vehicles discharging during the yellow indication).

$$G = C \frac{q_{in}}{q_s} \qquad (34)$$

where $q_s$ is the saturation flow rate and $C$ is the traffic signal cycle length.

Fig. 9 shows that the flow of vehicles when control is activated is, on average, lower than when it is not activated. Fig. 9 also shows that the controller attempts to maintain a constant inflow that targets the optimum operating point of ($\bar{k}, q_{max}$).

Fig. 10 shows the outflow of vehicles from the protected area when the control is not activated (NPC) and activated (SMC). It is important to note here that when the control is activated from the time step of 33 to 62, the outflow of vehicles from the protected area (in the control case) is higher than the NPC. This demonstrates that congested occurs in the protected area for NPC, which results in lower system throughput.

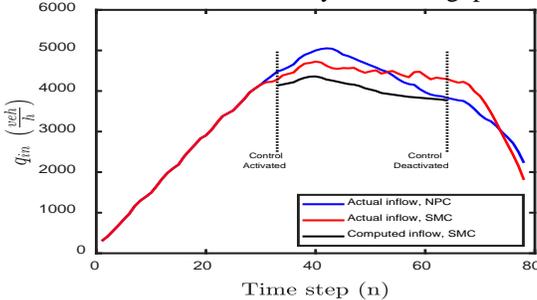

Fig. 9. Time series of the total vehicle inflow to the protected area from all the entry points with and without control as well as the computed control flow by the introduced logic.

### C. Performance of the SMC for Various Uncertainty Ranges on $q_{out}$ and $q_d$

Table 2 shows the effects of the uncertainty on $q_{out}$, $\alpha$ and on $q_d$, $\beta$. It is important to mention here that after careful examination of Equation (32), $\alpha$ and $\beta$ play exactly the same role in the controller.

As expected, as the uncertainty bound increases, the performance of the controller degrades (i.e., lower reduction in travel time, delay, fuel consumption, and lower increase in the average vehicular speed relative to the base case; Table 2).

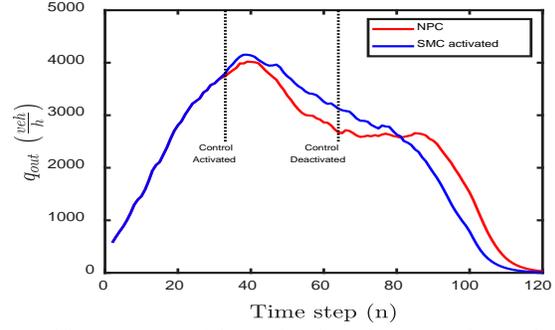

Fig. 10. Time series of the outflow from the protected area when the control is not activated and activated.

TABLE 2
SMC PERFORMANCE WITH RESPECT TO THE BASE CASE FOR VARIOUS
VALUES OF $\alpha$ AND $\beta$

| SMC | | | | | |
|---|---|---|---|---|---|
| $\alpha$ | $\beta$ | TT Change (%) | Delay Change (%) | Fuel Change (%) | Speed Change (%) |
| 25.00 | 0.00 | -12.75 | -17.57 | -6.13 | 12.07 |
| 50.00 | 0.00 | -10.10 | -13.37 | -4.84 | 8.94 |
| 100.00 | 0.00 | -7.65 | -9.27 | -3.10 | 6.93 |
| 300.00 | 0.00 | -8.10 | -4.04 | -1.53 | 7.96 |
| 400.00 | 0.00 | -7.49 | -5.51 | -2.07 | 6.86 |
| 0.00 | 25.00 | -12.75 | -17.57 | -6.13 | 12.07 |
| 0.00 | 50.00 | -10.10 | -13.37 | -4.84 | 8.94 |
| 0.00 | 100.00 | -7.65 | -9.27 | -3.10 | 6.93 |
| 0.00 | 300.00 | -8.10 | -4.04 | -1.53 | 7.96 |
| 0.00 | 400.00 | -7.49 | -5.51 | -2.07 | 6.86 |

### D. Response to Different Demand Profiles

To further demonstrate the effectiveness of the SMC, we considered other demand profiles for this study. These are shown in Fig. 11. Fig. 11(a) shows a dome-shaped demand (i.e., demand D2), and Fig. 11(b) shows a sinusoidal demand profile (i.e., demand D3). The evolution of the vehicle density in the protected area for the different demands D1, D2, and D3 is presented in Fig. 12, Fig. 13, and Fig. 14. It is important to note here that both controllers start from the $0.85\bar{k}$ threshold to regulate the density inside the protected area to values around $\bar{k}$. Table 3 shows the results for the PIC and SMC. Table 3 presents the change in travel time, delay, fuel consumption, and speed with respect to a base case for the PIC and SMC for the demand profiles D1, D2, and D3. For the demand profile D1, we notice that both controllers have quite similar performance, with one or the other exceeding slightly in one or two measures (i.e., for instance, the PIC is slightly better in travel time with respect to the SMC). For demand profile D2, we notice a slight advantage of the SMC with respect to the PIC. This advantage is clear in demand D3. The SMC outperforms the PIC by a relatively significant margin in the case of D3. This shows that the SMC adapts to a changing demand pattern whereas the PIC is less adaptable. It should be noted here that dynamic re-routing is activated. Therefore, the demands at the various traffic signals might vary.

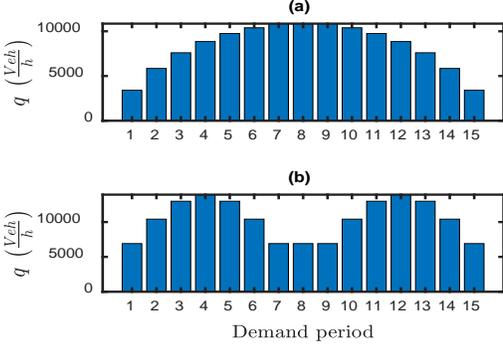

Fig. 11. Additional network demand profiles; each demand period spans 300 s. These demand profiles are referred to in the text as demand D2 and D3, respectively.

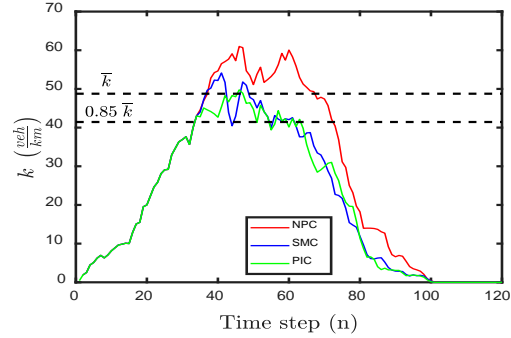

Fig. 12. Evolution of the density inside the protected area for the demand profile D1 presented in Fig. 4 without and with control.

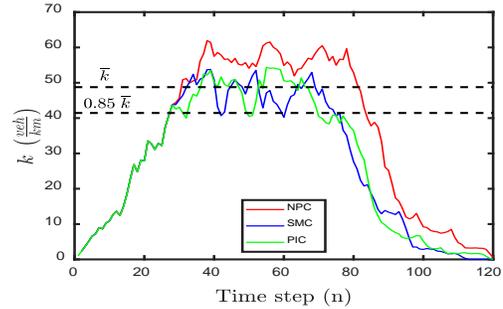

Fig. 13. Evolution of the density inside the protected area for the demand profile D2 presented in Fig. 11 (a) without and with control.

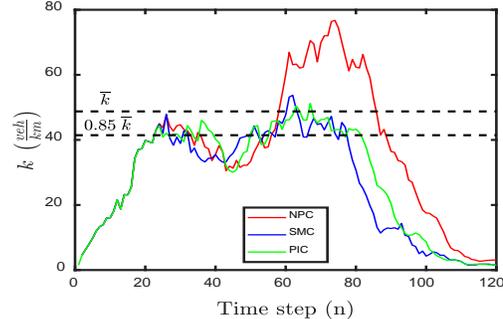

Fig. 14. Evolution of the density inside the protected area for the demand profile D3 presented in Fig. 11 (b) without and with control.

TABLE 3
AVG. SIMULATION RESULTS FOR THE DEMAND PROFILES OF FIG. 11

| | TT(s) | Delay (s) | Fuel (l) | Speed (km/h) |
|---|---|---|---|---|
| Demand Profile D1 | | | | |
| NPC | 659.26 | 247.28 | 0.42 | 14.97 |
| SMC Change (%) | -12.56 | -19.51 | -7.11 | 11.30 |
| PIC Change (%) | -12.95 | -17.52 | -6.25 | 12.12 |
| Demand Profile D2 | | | | |
| NPC | 956.11 | 401.66 | 0.51 | 10.73 |
| SMC Change (%) | -20.34 | -26.11 | -11.36 | 21.77 |
| PIC Change (%) | -18.76 | -27.04 | -11.12 | 20.93 |
| Demand Profile D3 | | | | |
| NPC | 950.76 | 415.29 | 0.52 | 11.02 |
| SMC Change (%) | -14.23 | -25.16 | -11.43 | 10.93 |
| PIC Change (%) | -8.28 | -17.78 | -7.37 | 6.11 |

## VII. CONCLUSION

In this paper, an SMC that attempts to alleviate traffic congestion in regions within large urban networks was developed. This controller was applied to the traffic stream flow continuity equation that governs the density of vehicles on any given link and thus a congested region. The controller was tested on a grid network and the results suggest that it has similar performance to the PIC and in some cases outperforms it. The SMC, however, does offer additional benefits, namely it makes no assumptions with regards to 1) the form or simplifications of the governing equations, 2) the existence or nonexistence of the NFD, or 3) the shape of the NFD (i.e., linearization of the NFD around the set point, then calibrating the PIC gains using the linearized data). This controller also has consistent performance with varying demand patterns. Unlike the PIC, it requires no tuning, but its parameters need be within a specified range. The user decides how fast the system should converge to the sliding surface, then once on the sliding surface how fast it converges to the desired control (error of zero). The controller consists of a single equation with various inputs that are collected from the detectors assumed to be available on all links within and feeding to the congested region. Future work will look into ways to reduce the number of detector measurements needed to run the controller.

## REFERENCES

[1] H. Rakha and M. Arafeh, "Calibrating steady-state traffic stream and car-following models using loop detector data," *Transportation Science*, vol. 44, no. 2, pp. 151-168, 2010.

[2] H. Rakha, "Validation of Van Aerde's Simplified Steady-state Car-following and Traffic Stream Model," *Transportation Letters: The International Journal of Transportation Research*, vol. 1(3), pp. 227-244, 2009.

[3] C. Buisson and C. Ladier, "Exploring the impact of homogeneity of traffic measurements on the existence of macroscopic fundamental diagrams," *Transportation Research Record: Journal of the Transportation Research Board*, no. 2124, pp. 127-136, 2009.

[4] Y. Ji, W. Daamen, S. Hoogendoorn, S. Hoogendoorn-Lanser, and X. Qian, "Investigating the shape of the macroscopic fundamental diagram using simulation data," *Transportation Research Record: Journal of the Transportation Research Board*, no. 2161, pp. 40-48, 2010.

[5] N. Geroliminis and J. Sun, "Properties of a well-defined macroscopic fundamental diagram for urban traffic," *Transportation Research Part B: Methodological*, vol. 45, no. 3, pp. 605-617, 2011.

[6] J. Godfrey, "The mechanism of a road network," *Traffic Engineering & Control*, vol. 8, no. 8, 1900.

[7] N. Geroliminis and C. F. Daganzo, "Existence of urban-scale macroscopic fundamental diagrams: Some experimental findings," *Transportation Research Part B: Methodological*, vol. 42, no. 9, pp. 759-770, 2008.

[8] L. Leclercq and N. Geroliminis, "Estimating MFDs in simple networks with route choice," *Procedia-Social and Behavioral Sciences*, vol. 80, pp. 99-118, 2013.

[9] C. F. Daganzo, "Urban gridlock: Macroscopic modeling and mitigation approaches," *Transportation Research Part B: Methodological*, vol. 41, no. 1, pp. 49-62, 2007.

[10] J. Haddad and A. Shraiber, "Robust perimeter control design for an urban region," *Transportation Research Part B: Methodological*, vol. 68, pp. 315-332, 2014.

[11] Y. Ji and N. Geroliminis, "On the spatial partitioning of urban transportation networks," *Transportation Research Part B: Methodological*, vol. 46, no. 10, pp. 1639-1656, 2012.

[12] M. Saeedmanesh and N. Geroliminis, "Clustering of heterogeneous networks with directional flows based on "Snake" similarities," *Transportation Research Part B: Methodological*, vol. 91, pp. 250-269, 2016.

[13] C. Lopez, P. Krishnakumari, L. Leclercq, N. Chiabaut, and H. Van Lint, "Spatiotemporal partitioning of transportation network using travel time data," *Transportation Research Record*, vol. 2623, no. 1, pp. 98-107, 2017.

[14] K. Aboudolas and N. Geroliminis, "Perimeter and boundary flow control in multi-reservoir heterogeneous networks," *Transportation Research Part B: Methodological*, vol. 55, pp. 265-281, 2013.

[15] J. Haddad and N. Geroliminis, "On the stability of traffic perimeter control in two-region urban cities," *Transportation Research Part B: Methodological*, vol. 46, no. 9, pp. 1159-1176, 2012.

[16] A. Kouvelas, M. Saeedmanesh, and N. Geroliminis, "A convex formulation for model predictive perime-ter flow control in multi-region cities," presented at the 16th Swiss Transport Research Conference (STRC 2016), 2016.

[17] C. Menelaou, S. Timotheou, P. Kolios, and C. Panayiotou, "Joint route guidance and demand management for multi-region traffic networks," in *2019 18th European Control Conference (ECC)*, 2019, pp. 2183-2188: IEEE.

[18] S. Batista and L. Leclercq, "Regional Dynamic Traffic Assignment Framework for Macroscopic Fundamental Diagram Multi-regions Models," *Transportation Science*, vol. 53, no. 6, pp. 1563-1590, 2019.

[19] R. Mohebifard and A. Hajbabaie, "Distributed optimization and coordination algorithms for dynamic traffic metering in urban street networks," *IEEE Transactions on Intelligent Transportation Systems*, pp. 1930-1941, 2018.

[20] K. Aboudolas, M. Papageorgiou, and E. Kosmatopoulos, "Store-and-forward based methods for the signal control problem in large-scale congested urban road networks," *Transportation Research Part C: Emerging Technologies*, vol. 17, no. 2, pp. 163-174, 2009.

[21] E. Christofa, K. Ampountolas, and A. Skabardonis, "Arterial traffic signal optimization: A person-based approach," *Transportation Research Part C: Emerging Technologies*, pp. 27-47, 2016.

[22] D. Manolis, T. Pappa, C. Diakaki, I. Papamichail, and M. Papageorgiou, "Centralised versus decentralised signal control of large-scale urban road networks in real time: a simulation study," *IET Intelligent Transport Systems*, vol. 12, no. 8, pp. 891-900, 2018.

[23] L. Bu, F. Wang, and X. Zhou, "Managed gating control strategy for emergency evacuation," *Transportmetrica A: Transport Science*, 2018.

[24] H. M. Abdelghaffar, H. Yang, and H. A. Rakha, "Isolated traffic signal control using a game theoretic framework," in *IEEE 19th International Conference on Intelligent Transportation Systems (ITSC)*, 2016.

[25] H. M. Abdelghaffar, H. Yang, and H. A. Rakha, "Developing a decentralized cycle-free nash bargaining arterial traffic signal controller," in *5th IEEE International Conference on Models and Technologies for Intelligent Transportation Systems (MT-ITS)*, 2017.

[26] R. Mohebifard, S. M. A. B. A. Islam, and A. Hajbabaie, "Cooperative traffic control and perimeter control in semi-connected urban-street networks," *Transportation Research Part C: Emerging Technologies*, vol. 104, pp. 408-427, 2019.

[27] R. Mohebifard and A. Hajbabaie, "Dynamic traffic metering in urban street networks: Formulation and solution algorithm," *Transportation Research Part C: Emerging Technologies*, vol. 93, pp. 161-178, 2018.

[28] R. Mohajerpoor, M. Saberi, H. L.Vu, T. M.Garoni, and M. Ramezani, "H∞ robust perimeter flow control in urban networks with partial information feedback," *Transportation Research Part B: Methodological*, 2019.

[29] N. Geroliminis, J. Haddad, and M. Ramezani, "Cooperative traffic control of a mixed network with two urban regions and a freeway," *Transportation Research Part B: Methodological*, vol. 54, pp. 17-36, 2013.

[30] J. Haddad and B. Mirkin, "Coordinated distributed adaptive perimeter control for large-scale urban road networks," *Transportation Research Part C: Emerging Technologies*, vol. 77, pp. 495-515, 2017.

[31] J. Haddad and Z. Zheng, "Adaptive perimeter control for multi-region accumulation-based models with state delays," *Transportation Research Part B: Methodological*, 2018.

[32] J. Haddad and B. Mirkin, "Adaptive perimeter traffic control of urban road networks based on MFD model with time delays," *International Journal of Robust and Nonlinear Control*, 2016.

[33] I. I. Sirmatel and N. Geroliminis, "Economic model predictive control of large-scale urban road networks via perimeter control and regional route guidance," *IEEE Transactions on Intelligent Transportation Systems*, vol. 19, no. 4, pp. 1112-1121, 2018.

[34] A. Kouvelas and N. Geroliminis, "Hierarchical control for large-scale urban road traffic networks," *7th International Conference on Transport Network Reliability*, 2018.

[35] J. Haddad, M. Ramezani, and N. Geroliminis, "Model predictive perimeter control for urban areas with macroscopic fundamental diagrams," *American Control Conference (ACC), 2012*, pp. 5757-5762, 2012.

[36] G. Mariotte and L. Leclercq, "Heterogeneous perimeter flow distributions and MFD-based traffic simulation," *Transportmetrica B: Transport Dynamics*, pp. 1-24, 2019.

[37] Y. Li, J. Xu, and L. Shen, "A perimeter control strategy for oversaturated network preventing queue spillback," *Procedia-Social and Behavioral Sciences*, vol. 43, pp. 418-427, 2012.

[38] M. Keyvan-Ekbatani, A. Kouvelas, I. Papamichail, and M. Papageorgiou, "Exploiting the fundamental diagram of urban networks for feedback-based gating," *Transportation Research Part B: Methodological*, vol. 46, no. 10, pp. 1393-1403, 2012.

[39] M. Elouni, H. A. Rakha, and Y. Bichiou, "Implementation and Investigation of a Weather-and Jam Density-Tuned Network Perimeter Controller," in *Smart Cities, Green Technologies and Intelligent Transport Systems*: Springer, 2018, pp. 266-278.

[40] M. Keyvan-Ekbatani, M. Papageorgiou, and V. L. Knoop, "Controller design for gating traffic control in presence of time-delay in urban road networks," *Transportation Research Procedia*, vol. 7, pp. 651-668, 2015.

[41] M. Keyvan-Ekbatani, X. Gao, V. V. Gayah, and V. L. Knoop, "Traffic-responsive signals combined with perimeter control: investigating the benefits," *Transportmetrica B: Transport Dynamics*, vol. 7, no. 1, pp. 1402-1425, 2019.

[42] T. A. M. Euzébio and P. R. Barros, "Optimal Integral Gain for Smooth PI Control," *IFAC Proceedings Volumes*, vol. 46, no. 11, pp. 529-533, 2013/01/01/ 2013.

[43] D. Singh, N. Singh, B. Singh, and S. Prakash, "Optimal gain tuning of PI current controller with parameter uncertainty in DC motor drive for speed control," in *2013 Students Conference on Engineering and Systems (SCES)*, 2013, pp. 1-6.

[44] R. Thangaraj, T. R. Chelliah, M. Pant, A. Abraham, and C. Grosan, "Optimal gain tuning of PI speed controller in induction motor drives using particle swarm optimization," *Logic Journal of the IGPL*, vol. 19, no. 2, pp. 343-356, 2011.

[45] R. Anandanatarajan, M. Chidambaram, and T. Jayasingh, "Limitations of a PI controller for a first-order nonlinear process with dead time," *ISA Transactions*, vol. 45, no. 2, pp. 185-199, 2006/04/01/ 2006.

[46] S. W. Sung and I-B. Lee, "Limitations and Countermeasures of PID Controllers," *Industrial & Engineering Chemistry Research*, vol. 35, no. 8, pp. 2596-2610, 1996/01/01 1996.

[47] M. Elouni and H. Rakha, "Weather-Tuned Network Perimeter Control-A Network Fundamental Diagram Feedback Controller Approach," in *VEHITS*, 2018, pp. 82-90.

[48]    A. Kouvelas, M. Saeedmanesh, and N. Geroliminis, "Enhancing model-based feedback perimeter control with data-driven online adaptive optimization," *Transportation Research Part B: Methodological,* vol. 96, pp. 26-45, 2017.

[49]    B. Mirkin, J. Haddad, and Y. Shtessel, "Tracking with asymptotic sliding mode and adaptive input delay effect compensation of nonlinearly perturbed delayed systems applied to traffic feedback control," *International Journal of Control* vol. 89, no. 9, 2016.

[50]    C. Edwards and S. Spurgeon, *Sliding mode control: theory and applications.* Crc Press, 1998.

[51]    J.-J. E. Slotine and W. Li, *Applied nonlinear control.* Prentice hall, 1991.

[52]    V. I. Utkin and A. S. Poznyak, *Adaptive Sliding Mode Control* (Advances in Sliding Mode Control. Lecture Notes in Control and Information Sciences). 2013.

[53]    F. Dion, H. Rakha, and Y.-S. Kang, "Comparison of delay estimates at under-saturated and over-saturated pre-timed signalized intersections," *Transportation Research Part B-Methodological,* vol. 38, no. 2, pp. 99-122, 2004.

[54]    H. Rakha, Y. S. Kang, and F. Dion, "Estimating vehicle stops at undersaturated and oversaturated fixed-time signalized intersections," (in English), *Traffic Flow Theory and Highway Capacity 2001,* no. 1776, pp. 128-137, 2001.

[55]    H. Rakha, Y. Huanyu, and F. Dion, "VT-Meso model framework for estimating hot-stabilized light-duty vehicle fuel consumption and emission rates," *Canadian Journal of Civil Engineering,* vol. 38, no. 11, pp. 1274-86, 2011.

[56]    H. Rakha, K. Ahn, and A. Trani, "Development of VT-Micro model for estimating hot stabilized light duty vehicle and truck emissions," *Transportation Research Part D: Transport and Environment,* vol. 9, no. 1, pp. 49-74, 2004.

[57]    K. Ahn, H. Rakha, A. Trani, and M. Van Aerde, "Estimating vehicle fuel consumption and emissions based on instantaneous speed and acceleration levels," *Journal of Transportation Engineering,* vol. 128, no. 2, pp. 182-190, 2002.

[58]    R. P. Roess, E. S. Prassas, and W. R. McShane, *Traffic Engineering.* Pearson, 2010.

[59]    H. Rakha and M. V. Aerde, "An Off-line Emulator for Estimating the Impacts of SCOOT," presented at the 74th Transportation Research Board Annual Meeting, Washington DC, 1995.

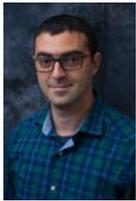

**Youssef Bichiou** received his Ph.D. in Engineering Mechanics from Virginia Polytechnic Institute and State University (Virginia Tech). He finished his graduate studies in the field of fluid structure interaction in the department of Mechanical Engineering, Tunisia Polytechnic School, La Marsa, Tunisia. He is a Research Associate in the Center for Sustainable Mobility at the Virginia Tech Transportation Institute, Blacksburg, Virginia. His research interests include traffic flow theory, autonomous vehicle control, traffic congestion alleviation, and intersection control.

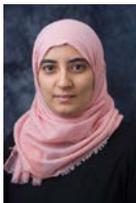

**Maha Elouni** received her B.Sc. and M.Sc. in computer science from the National School of Computer Science in Tunisia, both in 2012, and her M.Sc. in applied mathematics from Virginia Tech in 2015. She is currently a Ph.D. student in the Bradley Department of Electrical and Computer Engineering at Virginia Tech and works as a Graduate Research Assistant in the Center for Sustainable Mobility at the Virginia Tech Transportation Institute. Her research interests include traffic control, traffic modeling and simulation, and intelligent transportation systems.

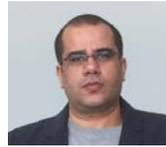

**Hossam M. Abdelghaffar** received his B.Sc. (with honors) in Electronics Engineering from the Faculty of Engineering, Mansoura University, Egypt, his M.Sc. in Automatic Control System Engineering from Mansoura University, Egypt, and his Ph.D. in Electrical Engineering from the Bradley Department of Electrical and Computer Engineering, Virginia Tech. He is currently an Assistant Professor in the Department of Computer Engineering and Systems, Faculty of Engineering, Mansoura University, Egypt, and with the Center for Sustainable Mobility at the Virginia Tech Transportation Institute.

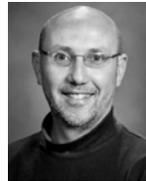

**Hesham Rakha** (M'04, SM'18, F'20) received his B.Sc. (with honors) in civil engineering from Cairo University, Cairo, Egypt, in 1987 and his M.Sc. and Ph.D. in civil and environmental engineering from Queen's University, Kingston, ON, Canada, in 1990 and 1993, respectively. Dr. Rakha's research focuses on large-scale transportation system optimization, modeling and assessment. Specifically, Dr. Rakha and his team have expanded the domain of knowledge (in traveler and driver behavior modeling) and developed a suite of multi-modal agent-based transportation modeling tools, including the INTEGRATION microscopic traffic simulation software. This software was used to evaluate the first dynamic route guidance system, TravTek in Orlando, Florida; to model the Greater Salt Lake City area in preparation for the 2002 Winter Olympic Games; to model sections of Beijing in preparation for the 2008 Summer Olympic Games; to optimize and evaluate the performance of alternative traveler incentive strategies to reduce network-wide energy consumption in the Greater Los Angeles area; and to develop and test an Eco-Cooperative Automated Control (Eco-CAC) system. Finally, Dr. Rakha and his team have developed various vehicle energy and fuel consumption models that are used world-wide to assess the energy and environmental impacts of ITS applications and emerging Connected Automated Vehicle (CAV) systems. The models include the VT-Micro, VT-Meso, the Virginia Tech Comprehensive Fuel consumption Model (VT-CPFM), the VT-CPEM, and the VT-CPHEM models.